 \documentclass[final,5p,times]{elsarticle}

\usepackage{amssymb}

\usepackage{lineno}

\usepackage{graphicx}
\usepackage{caption}
\usepackage{subcaption}
\usepackage{float}
\usepackage[thinc]{esdiff}
\usepackage{amsmath}
\usepackage{nicefrac}
\usepackage{mathtools}
\usepackage{dcolumn}
\usepackage{bm}
\usepackage{widetext}
\usepackage{color,soul}
\usepackage{caption}

\usepackage{hyperref}

\hypersetup{
    colorlinks=false,
    linkcolor=blue,
    filecolor=magenta,      
    urlcolor=cyan,
    pdftitle={Overleaf Example},
    pdfpagemode=FullScreen,
    }
    
\usepackage{tabularray}

\begin{document}

\begin{frontmatter}



\title{Optical and acoustic ground effects simulations from terminal defense asteroid disruption via the PI method}


\author[label1]{Brin Bailey}
\ead{brittanybailey@ucsb.edu}
\author[label1]{Alexander N. Cohen}
\author[label1]{Philip Lubin}
\author[label2]{Darrel Robertson}
\author[label3]{Mark Boslough}
\author[label4]{Sasha Egan}
\author[label5]{Elizabeth A. Silber}
\author[label1]{Dharv Patel}

\affiliation[label1]{organization={University of California - Santa Barbara},
            addressline={Broida Hall}, 
            city={Santa Barbara},
            state={California},
            postcode={93106}, 
            country={United States}}
\affiliation[label2]{organization={NASA Ames Research Center},
            city={Moffett Field},
            state={California},
            postcode={94035},             
            country={United States}}
\affiliation[label3]{organization={University of New Mexico},
            addressline={1700 Lomas Blvd, NE. Suite 2200}, 
            city={Albuquerque},
            state={New Mexico},
            postcode={87131},
            country={United States}}
\affiliation[label4]{organization={New Mexico Institute of Mining and Technology, Energetic Materials Research and Testing Center},
            city={Socorro},
            state={New Mexico},
            postcode={87801},
            country={United States}}
\affiliation[label5]{organization={Sandia National Laboratories},
            city={Albuquerque},
            state={New Mexico},
            postcode={87123},
            country={United States}}

\begin{abstract}
Our simulations suggest that PI (``Pulverize It"), a NASA Phase II NIAC study, is an effective multi-modal approach for planetary defense that can operate in extremely short interdiction modes (with intercepts as short as hours prior to atmospheric entry) as well as long interdiction time scales with months to years of warning. The basic process is complete disruption of the threat via fragmentation. In scenarios with sufficiently long warning time, the fragment cloud spreads enough to miss Earth, resulting in no ground effects. In ``worst-case" scenarios, when the warning time is short, the fragments (typically \textless10 m in diameter) will enter Earth's atmosphere, where their energy is dissipated in a series of ground-level optical pulses and de-correlated shock waves, mitigating any significant damage. We investigate the optical and acoustic ground effects through a set of simulation codes that model the interaction of asteroid fragments with Earth's atmosphere following terminal threat interception. Even in short-warning time cases where fragments enter the atmosphere, our simulations suggest that threats mitigated by the PI method produce vastly less damage on the ground when compared to the same unfragmented case, yielding optical energy deposition below 200 kJ/m$^2$ and shock wave over-pressures under 3 kPa. Our simulations support the proposition that threats like 2023 PDC, the hypothetical 800 m diameter asteroid from the 2023 Planetary Defense Conference impact exercise, can be effectively mitigated through fragmentation. We find that a terminal defense mitigation scenario that disrupts 2023 PDC into 1 million fragments with an intercept of 60 days before ground impact results in minimal ground effects.
\end{abstract}




\end{frontmatter}


\section{\label{sec:intro}Introduction}

\subsection{\label{sec:impacts}Earth impact threats}
When asteroids enter Earth's atmosphere, they either explode in the air (airburst) or impact the ground. Both processes emit shock waves, blast winds, and thermal radiation (collectively referred to here as ``ground effects") that have the capacity to enact damage far beyond the entry trajectory \cite{robertson_hydrocode_2019}. The majority of asteroids are small in diameter; there is an estimated population of about 2E+09 Near-Earth Asteroids (NEAs) of diameter $\leq100$ m compared to only about 940 NEAs $\geq1$ km in diameter \cite{harris_population_2021}. While small asteroid airbursts are far less destructive than impacts of larger threats, historical events have shown that small asteroids can pose significant hazards. 

The Chelyabinsk airburst event in 2013 (which released energy equivalent to $0.57\pm0.15$ Mt TNT) and the Tunguska event in 1908 (which produced an estimated yield of 3-15 Mt) caused disruptions to local human life and land \cite{popova_chelyabinsk_2013,brown_flux_2002,jenniskens_tunguska_2019,longo_tunguska_2007}. Objects at least the size of the Chelyabinsk asteroid (18 m diameter) are expected to impact Earth approximately every 50-100 years, while objects similar to the Tunguska asteroid (estimated $\sim$50 m diameter) are expected to impact Earth approximately every 300-1800 years ($\sim$300-500 years for an estimated 3-5 Mt event; $\sim$800-1800 years for an estimated 10-15 Mt event) \cite{harris_population_2021,boslough_low-altitude_2008,brown_flux_2002}. Understanding the ground damage caused by small asteroid airbursts is valuable in determining the proper response to a potential impact, such as evacuation or mitigation via planetary defense.

While small asteroids are the most common, larger objects maintain the possibility of posing a great threat to human life. The near-Earth asteroid 99942 Apophis, a potentially hazardous asteroid (PHA) with an equivalent diameter of $\sim350$ m, will make its next closest approach to Earth on April 13, 2029, where it will approach within Earth’s geosynchronous orbit. A PHA of this size encountering Earth this closely is expected to occur approximately every 1000 years \cite{binzel_apophis_2021}. For reference, Apophis is 7 times larger and 350 times more massive than the Tunguska 1908 impactor \cite{robertson_hydrocode_2019}, while 5000 times more massive than the Chelyabinsk 2013 asteroid \cite{popova_chelyabinsk_2013}. If Apophis were to collide with Earth, it would have an impact yield of approximately 3-4 Gt TNT equivalent \cite{brown_flux_2002}, equal to about half of Earth’s total nuclear arsenal. However, note that we introduce Apophis only for comparison and not as a viable threat, as continuous observations of Apophis since its discovery in 2004 have ruled out the possibility of a collision with Earth for at least the next 100 years \cite{noauthor_nasa_2021}. Larger than Apophis is asteroid 101955 Bennu with a diameter of $\sim$500 m whose impact could pose a yield comparable to Earth's entire nuclear arsenal \cite{brown_flux_2002}. The Jet Propulsion Laboratory (JPL) Sentry System estimates several potential impact scenarios of Bennu \cite{noauthor_sentry_2023} (as of April 15, 2024); however, each estimated probability is so low as to be effectively zero. While these particular objects currently pose no threat to Earth, it is conceivable that objects of similar sizes could potentially impact Earth in the future. 

To achieve an extensive planetary defense system, preparedness for a variety of threat scenarios–considering a large range of threat sizes and warning times–is imperative. It can be argued that a robust planetary defense system would be comprised of a layered system of reliable, tested methods for both detection and mitigation to achieve such preparedness. While we focus on mitigation in this study, it is important to note that a distinct constraint to mitigation is the necessity to observe a threat prior to its potential impact; detection plays a crucial factor in planetary defense. We propose PI (Section \ref{sec:PI}) as a potential multi-modal mitigation method not to replace detection or other mitigation methods, but as a supplementary approach to bolster our response to potential threats.

\subsection{\label{sec:PI}PI for planetary defense}
PI (``Pulverize It") is a NASA Phase II NIAC (NASA Innovative Advanced Concepts) study of planetary defense which is intended to operate in both terminal and extended interdiction modes, representing a fundamentally different approach to threat mitigation. Planetary defense has traditionally focused on mitigation via orbital modification or deflection, utilizing momentum transfer to prevent an impact. Deflection is applied in a range of techniques, from impulsive methods like direct impact \cite{rivkin_double_2021} or nuclear ablation \cite{dearborn_options_2020}, to gradual orbit deflection (e.g., via surface albedo alteration) \cite{hyland_permanently-acting_2010}, or to the utilization of gravity tractors \cite{mazanek_enhanced_2015}, ion engines, laser ablation, and further technologies \cite{walker_concepts_2005}. PI is an alternative approach which uses energy transfer for mitigation. The method utilizes an array of hypervelocity kinetic penetrators that disassemble an asteroid into many small (typically \textless10 m) fragments \cite{lubin_asteroid_2023}. Options for explosive penetrators including nuclear explosive devices (NEDs) are also a part of our study.

PI represents a multi-modal planetary defense capability: depending on the time scale of interception, the fragment cloud either misses Earth entirely (long-warning time) or is dissipated in Earth’s atmosphere (short-warning time, henceforth referred to as the “terminal mode”) \cite{lubin_asteroid_2023}. The latter results in a series of airburst events with spatial and temporal spread at varying high altitudes which distributes the energy of the parent asteroid \cite{lubin_asteroid_2023}.

We must note that, while effective in short-warning scenarios, PI's preferred usage is in extended mitigation scenarios with interception on the order of months to years, so as to prevent interaction between the fragment cloud and Earth. This paper focuses on use of PI in its terminal defense mode to analyze its effectiveness in "worst-case" scenarios, with a particular emphasis on short-warning mitigation of the hypothetical threat scenario 2023 PDC (Section \ref{sec:800m}). Further description of PI's multi-modal capability in extended interdiction modes can be found in Lubin and Cohen \cite{lubin_asteroid_2023}.

\subsection{\label{sec:terminal defense}Terminal planetary defense using PI}
In the terminal mode (hours-to-days intercept; 15–100 m diameter threats), the impacting fragment cloud interacts with Earth's atmosphere in a similar manner to an unmitigated asteroid airburst, but instead disperses the energy relative to the unmitigated case \cite{lubin_asteroid_2023}. During atmospheric entry of the fragments, the high-speed ram pressure (or stagnation pressure) exerted by the atmosphere eventually exceeds the material yield strength, initiating a cascading breakup event \cite{kring_chelyabinsk_2014,robertson_effect_2017}. The ram pressure is determined by the density of the atmosphere and speed of the parent asteroid, whereas yield strength depends largely on the shear strength provided by the internal structure and integrity of the asteroid, including the strength of its components (as well as other parameters such as size, density, speed, and entry angle) \cite{robertson_effect_2017}. 

As the pressure buildup on the fragment increases, it undergoes ablation, causing outward expansion of material and thereby increasing the surface area on which the rising aerodynamic drag can act \cite{kring_chelyabinsk_2014}. This runaway process eventually converts the fragment’s kinetic energy into a release of heat and pressure through detonation, or ``bursting," of the fragment \cite{kring_chelyabinsk_2014}. Depending on the material strength, initial failure can occur either externally or internally; the failure site will influence the method by which the fragment bursts \cite{robertson_effect_2017}. 

These airbursts yield optical pulses and de-correlated shock waves on the ground (hereafter referred to as ``ground effects") which, in reasonable mitigation scenarios that are appropriate for the threat, result in little to no damage.

\subsection{\label{sec:ground effects}Ground effects}
The ground effects of unmitigated asteroids or very large ($\textgreater$20 m) fragments have the potential for significant destruction; we therefore design mitigation scenarios such that our fragments are generally $\textless$10 m in diameter \cite{lubin_asteroid_2023}. It is important to analyze both the optical and acoustic ground effects to design proper mitigation scenarios with acceptably low damage.


The maximum permissible exposure (MPE) to optical damage can be characterized in both total energy deposition (J/$\text{m}^2$) and time-dependent power output (W/$\text{m}^2$). High energy deposition can lead to hazards such as fires, skin damage (sunburn), and retinal damage \cite{glasstone_effects_1977}. We set an optical energy damage threshold of 200 kJ/$\text{m}^2$ such that the sum of the optical energy over all fragments (assuming the analytic relationship outlined in Section \ref{sec:optical methods}) is kept at or below this value. We choose a value of 200 kJ/$\text{m}^2$ ($\sim$5 cal/cm$^2$ of radiant exposure) as this is the point at which combustible organic materials (like leaves and paper) can begin to catch fire \cite{martin_diffusion-controlled_1965}. 

In regard to optical power output, we make the assumption that the optical power deposition from fragment bursts can be approximated by those of atmospheric nuclear tests. The occurrence of biological hazards such as flashblindness or retinal burns depends on a variety of factors, namely explosion yield, height of burst, observer distance from ground zero, exposure time, weather (clear skies versus cloud cover), and time of day (day versus night) \cite{glasstone_effects_1977}. It is therefore difficult to establish one consistent damage threshold for optical power output. It seems a safe assumption that, in a hypothetical mitigation scenario (particularly those with short warning time), the best and most consistent protection from optical damage is to stay indoors or shield one's eyes. In a hypothetical terminal defense scenario that would necessitate shielding of the eyes, it seems reasonable to assume that preventative civil defense measures could be made to mitigate potential damage to the public, comparable to the alerts given before a solar eclipse.

For acoustic damage, studies of window damage in atmospheric nuclear tests \cite{glasstone_effects_1977} have found that the threshold for residential window breakage corresponds to peak pressures of about 3 kPa. We then set a shock wave over-pressure threshold of 3 kPa such that our goal for any mitigation scenario is to keep all shock wave over-pressures (including interference) below this value.

\section{\label{sec:optical methods}Optical pulse modeling}
The conversion of kinetic energy into optical energy is highly dependent on fragment properties, particularly cohesive strength, and is poorly understood in general. We resort to measured optical data to model this conversion, primarily from Department of Defense satellite observations of a small number of relevant bolide sizes of interest to us (typically 1–10 m diameter) \cite{brown_flux_2002}.

Using an analytical extrapolation from \cite{brown_flux_2002}, we calculate the optical energy at burst (in Joules) from exo-atmospheric energy $E_{exo}$ as
\begin{equation} \label{eq.analytical op}
    E_{opt} = \left( \frac{E_{exo}}{8.2508} \right) ^{1.13}
\end{equation}

For the propagation of the optical pulse through the atmosphere, we use a full radiation transfer model to compute the optical power flux from each fragment at each observer. Because the optical propagation is occurring at very close to the speed of light in vacuum and the relevant distance scale from the fragment to the observer are of order of tens to hundreds of km, the light propagation time scale is very short. The optical pulse can then be well approximated as happening nearly simultaneously at all observer points, with the optical pulse arriving very shortly after the fragment burst. 

As a result, we propagate the optical energy flux to each observer using the distance propagation of light, inputting $E_{opt}$ at the time of burst to simulate an instantaneous addition of energy flux \cite{lubin_asteroid_2023}. We propagate the optical emission from every fragment to the observer to get a total energy flux at the observer.

\subsection{\label{sec:blackbody}Atmospheric attenuation and cooling}
Several factors can greatly affect the observed optical flux, such as the source spectral energy distribution and atmospheric attenuation, which depend on both the source and the complex and time-varying nature of the atmosphere \cite{glasstone_effects_1977}. To calculate the attenuation of the optical signature, we approximate the parent asteroid as a blackbody source due to the wavelength-dependent transmission of the atmosphere \cite{lubin_asteroid_2023}. We model the atmosphere using MODTRAN to perform a full analysis of attenuation which includes the curvature of Earth’s atmosphere and the relationships between nominal atmospheric pressure, temperature, and altitude \cite{lubin_asteroid_2023}. 

Note that we assume an extremely conservative case of no cooling at the observer between fragment optical pulses. However, in a real scenario, fragments arrive on the order of tens of seconds apart (hundreds of seconds in some extreme cases) which would, in general, be enough time for significant cooling between bursts of incident optical energy. This will be important when judging the effectiveness of this method if the energy exceeds our threshold.

\subsection{\label{sec:optical vis}Optical pulse visualization}
Optical light curves from satellite observations of small bolides (1-10 m size range) \cite{brown_flux_2002,chamberlin_fireballs_2024} can generally be well-approximated by a Gaussian fit with time dispersion ($\sigma$) of order 0.1-1 s (depending on fragment parameters, particularly size). As a result, our optical pulse visualizations (Figure \ref{fig:800m GE}, left) utilize a Gaussian distribution to describe the power as a function of time produced by each fragment, typically assuming a time duration of each light pulse as $\sigma\approx0.1-1$ s. Note that this approximation is purely for aesthetics in visually displaying the optical pulses and does not affect our calculations of optical energy deposition. As our model conservatively assumes no cooling between optical pulses, the optical energy flux at the observer is calculated as the sum of the optical energy flux from each pulse.

\section{\label{sec:acoustic methods}Acoustic shock wave modeling}
The acoustic ground effects from an asteroid fragment airburst can be related to and approximated by those of nuclear blasts as discussed by Boslough et al. \cite{boslough_updated_2015}. As such, we base our simulations of the acoustic ground effects from mitigation via PI on measurements of equivalent nuclear blasts. To model the time evolution of the shock wave, we use a Friedlander functional form, given by
\begin{equation}\label{time evolution}
    p(t,r)=p_0(r)e^{-t/t_1}(1-t/t_1)
\end{equation}
which describes the shock wave time evolution at a distance $r$ with two free parameters: the peak pressure $p_0$ in Pa at time $t=0$ seconds and the Friedlander positive pulse time scale, or zero crossing time, $t_1$ in seconds. Note that the time $t=0$ seconds is the time at which the shock wave first arrives at the observer, not to be confused with the time at which the fragment bursts.

Letting $\epsilon$ denote the fraction of a 1 kt yield that goes into the shock wave (typically 0.5), we achieve $E_{nuc}=E_{ast-kt}/\epsilon$, where $E_{ast-kt}$ is the asteroid airburst shock wave energy and $E_{nuc}$ is the equivalent energy of a nuclear weapon \cite{lubin_asteroid_2023}. The peak pressure at a distance $r$ is calculated from the equivalent energy of a nuclear weapon as
\begin{equation}\label{peak pressure}
    p(r)=p_n{[rE_{nuc}^{1/3}]}^{\alpha_n}+p_f{[rE_{nuc}^{1/3}]}^{\alpha_f}
\end{equation}
where $p_n=3.11\times10^{11}$ Pa is the pressure for a 1 kt standard weapon yield in the near field, $\alpha_n=-2.95$ is the power law index for the near field, $p_f=1.80\times10^{7}$ Pa is the pressure for a 1 kt standard weapon yield in the far field, and $\alpha_f=-1.13$ is the power law index for the far field \cite{lubin_asteroid_2023}.

We simulate the airburst produced by each fragment as it enters Earth's atmosphere using equations (\ref{time evolution}) and (\ref{peak pressure}). The model considers any interference between interacting shock waves, summing them to simulate the acoustic caustics.

\subsection{\label{sec:caustics}Shock wave caustics}
As a fragment airbursts, the emitted shock wave can be thought of as an expanding sphere whose intersection with the ground plane forms a circle. Acoustic caustics form when shock waves from multiple fragment airbursts constructively interfere in the ground plane. Our model takes into account such interference to simulate the acoustic caustics that form as shock waves interact \cite{lubin_asteroid_2023}. Areas of constructive interference experience higher over-pressures, and thus they must be taken into account in order to design mitigation scenarios with acceptably low pressure values.

\subsection{\label{sec:t1}Friedlander positive pulse time scale, $t_1$}
The positive pressure shock wave duration $t_1$ in the Friedlander parametrization of the shock wave time evolution is dependent on a number of parameters, including blast yield, altitude of the blast source, distance from the blast, and atmospheric absorption. Note that in the case of nuclear airbursts, $\sim1/2$ of the total energy goes into the shock wave and related atmospheric effects (i.e., wind production) \cite{glasstone_effects_1977,brode_review_1968}. Thus, we roughly double the shock wave yield of an asteroid fragment to achieve the equivalent nuclear yield that would produce the same shock wave \cite{lubin_asteroid_2023}.

Data from both conventional and nuclear weapons tests and from measured airbursts show evidence of pulse stretching at large distances \cite{glasstone_effects_1977,brode_review_1968}. The positive pressure time $t_1$ then has a pressure-distance dependence: the pulse stretches with increased distance, thus decreasing over-pressure \cite{glasstone_effects_1977,brode_review_1968}. Thus, in the nuclear airburst tests from which this relation for $t_1$ is derived, we observe a relationship between peak pressure and $t_1$. However, we find that our simulations are not particularly sensitive to the $t_1$ value \cite{lubin_asteroid_2023}. As a result, we extrapolate the best fit from the regime of 10-100 kPa where there is applicable data \cite{glasstone_effects_1977,brode_review_1968} and assume the trend continues to lower over-pressures \cite{lubin_asteroid_2023}. 

We then scale the $t_1$ measurement from Brode \cite{lubin_asteroid_2023,brode_review_1968} to calculate the $t_1$ parameter using a relation to pressure. Our $t_1$ parameter, in seconds, is then
\begin{equation}
    t_1=
    \begin{cases}
        [-0.07755{\cdot}{\ln(p)}+1.051]*E_{kt}^{1/3} & \text{for $p<200$ kPa}
       \cr [0.01246{\cdot}{\ln(p)}-0.07758]*E_{kt}^{1/3} & \text{for $p>200$ kPa}
    \end{cases}
\end{equation}
where $p$ is in Pa and $E_{kt}$ is the total effective shock wave energy (including coupling coefficient between shock wave and yield, typically 0.5) in kt \cite{lubin_asteroid_2023}. While our extrapolation includes a measurement for $t_1$ at pressures above 200 kPa, for our purposes in calculating the acoustic ground effects we remain in the weak shock regime and typically utilize only the $t_1$ value for $p<200$ kPa \cite{lubin_asteroid_2023}.

\begin{figure}[ht]
    \centering
    \includegraphics[width=0.48\textwidth]{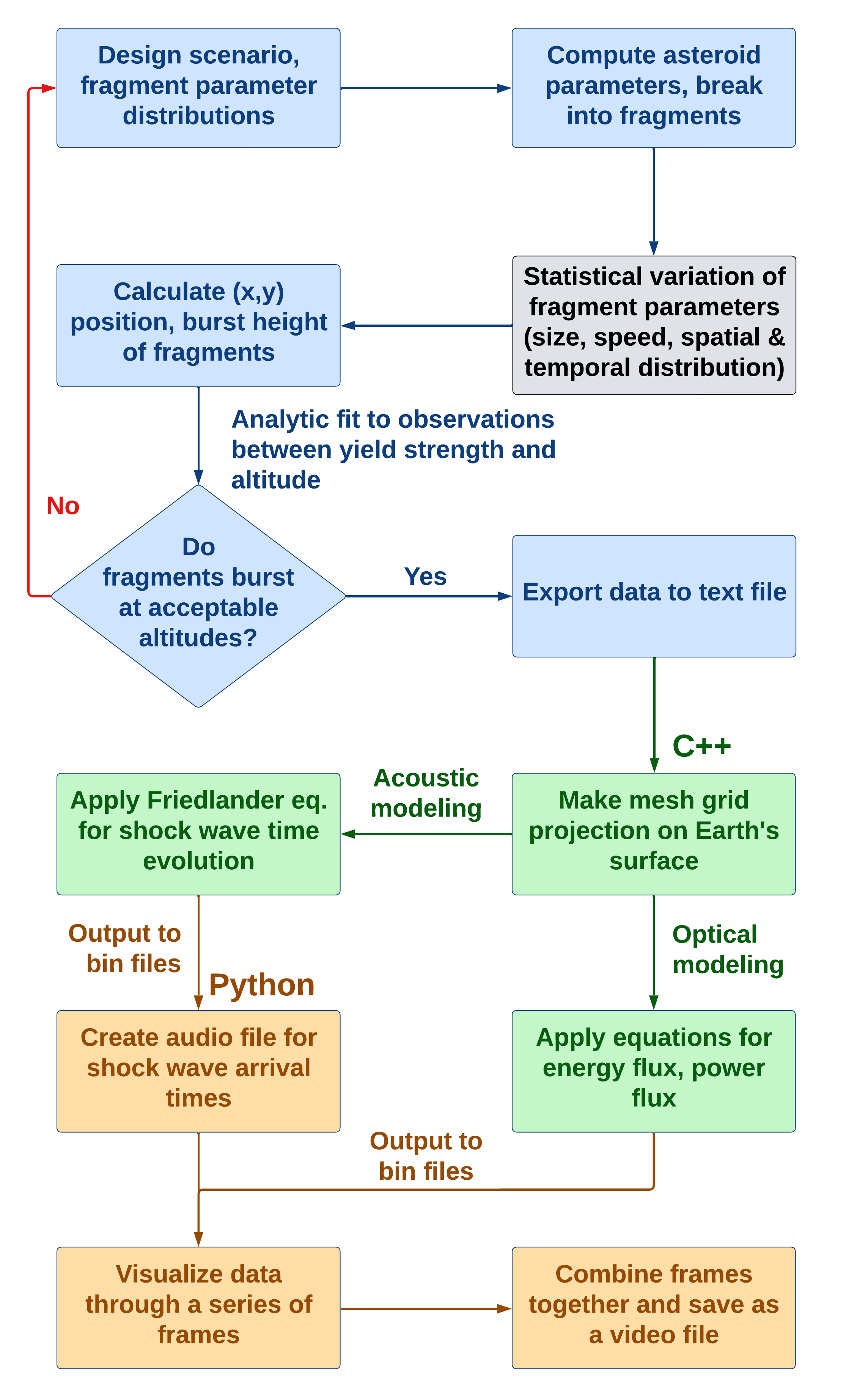}
    \caption{Process of parameter definition, computation, and visualization via custom codes to produce ground effects simulations.}
    \label{fig:flowchart}
\end{figure}

\section{\label{sec:codes}Ground effects codes}
We simulate the ground effects resulting from the fragments entering Earth's atmosphere for the terminal mode relevant in this paper via computation of the optical pulse model and analytic airburst model for a given scenario (Figure \ref{fig:flowchart}). As mentioned in Section \ref{sec:ground effects}, each scenario is designed to keep the ground effects (as observed by an arbitrary observer on Earth's surface) below their respective damage thresholds of 200 kJ/$\text{m}^2$ and 3 kPa in order to minimize ground damage. The magnitude of the effects for any given case is dependent upon the mitigation parameters of fragment number and intercept time (the time between intercept and ground impact). In general, increasing the number of fragments and/or intercept time will decrease the magnitude of the ground effects \cite{lubin_asteroid_2023}; however, we find that exceptions exist, particularly in very short intercept scenarios ($\le$1 day) (Figure \ref{fig:20m 50m}).

Upon choosing a scenario to simulate, the input parameters are defined and ground effects are calculated. The primary inputs for the model include the parent asteroid diameter, speed, average density, average yield strength, entry angle, and number of fragments. All scenarios assume a spherical target body.

We introduce statistical variations in the fragmentation process to simulate the uncertainty in a real scenario for several fragment parameters, including diameter (fragment size), density, yield strength, burst location, slant distance away from airburst, and fragment disruption velocity (the asymptotic velocity at which fragments move away from the fragment cloud's center of mass, after having been decelerated by the self-gravity of the asteroid). Note that variation of these parameters is only of interest in terminal defense scenarios where fragments enter Earth's atmosphere.

For every fragment in a given case, each parameter is varied by a normal distribution with randomized probability. In a typical simulation, we set the standard deviation ($\sigma$) of each parameter to be $\sim10-30\%$ of the mean value; e.g., in varying fragment diameter $L_0$, we use $\sigma_L=L_{avg}*L_{disp}$ where $L_{avg}$ is the average fragment diameter and $L_{disp}$ is the fragment diameter dispersion, typically set as $L_{disp}=0.25L_{avg}$ \cite{lubin_asteroid_2023}. Additionally, we place limitations (minimum and/or maximum values) on several parameters to maintain a reasonable range of values; e.g., we set $L_0\geq2$ m, as small fragments are largely irrelevant in terms of ground effects.

The values we utilize for parameter means, ranges, and $\sigma$ values represent our best estimates for reasonable asteroid mitigation scenarios based upon 1) the very limited amount of available data regarding asteroid interiors and 2) the very wide breadth of asteroid interception and ground effects simulations we have carried out (the inclusion of which is beyond the scope of this paper). These values are not fine-tuned, nor are they designed to work with our simulations; rather, our simulations have been designed to investigate a relatively broad range of reasonable values.

For example, we have explored densities from 1-8 g/cm$^3$ and fragment yield strengths from several kPa to 500 MPa, the latter being extremely conservative \cite{lubin_asteroid_2023}. In all cases, we find that mitigation via fragmentation is possible, with smaller fragment sizes ($<$10 m diameter) always being desirable. We find that for typical rocky asteroid densities of 2-3 g/cm$^3$, fragment sizes less than $\sim$10-15 m are acceptable, while for cases with exceptionally high density and high yield strength, fragment sizes less than $\sim5$ m are typically acceptable \cite{lubin_asteroid_2023}.

For fragment disruption velocity ($v_{disr}$), we set a nominal value of 1 m/s and $\sigma_{v_{disr}}=0.3$ m/s, though we have explored cases with $v_{disr}$ up to 10 m/s \cite{lubin_asteroid_2023}. It is worth noting that the fragment cloud size $R$ is dependent upon the intercept time ($\tau$) and $v_{disr}$ as $R_{avg}=v_{disr_{avg}}*\tau$ (where $\tau$ is in seconds), so $R{\propto}v_{disr}$ \cite{lubin_asteroid_2023}. As this study considers cases in which the fragment cloud intercepts Earth, all simulations presented here (Section \ref{sec:sims}) set the average fragment disruption velocity to the nominal value.

\subsection{\label{sec:c++}Fragment cloud projection}
We then input the data into a C++ code which forms a surface map of the fragment cloud of any given size for any number of fragments, with a current upper limit of 1 million fragments. A 3D array is created to form a mesh grid projection of the fragment cloud, mapping out the spatial and temporal spread across a specified number of frames. For optical modeling, we apply the Gaussian waveform to visualize the optical energy and optical power of all bursts within the simulation area and time frame. For acoustic modeling, we apply the Friedlander functional form to produce the shock wave time evolution of all bursts within the simulation area and time frame.

\subsection{\label{sec:python}Data visualization}
For each model, a Python code reads the C++ output and utilizes a plotting library to visualize the data into five separate plots. For optical modeling, we achieve real-time simulations of optical power flux, optical energy flux, optical power flux distribution, and maximum optical power, as well as a cumulative distribution function (CDF) which describes the frequency of occurrence of various optical values \cite{lubin_asteroid_2023}. For acoustic modeling, we achieve real-time simulations of pressure, maximum pressure, pressure distribution, and maximum and minimum pressure, as well as a CDF displaying the frequency of occurrence of pressures \cite{lubin_asteroid_2023}. 

An additional input for acoustic modeling is the position (in x and y) of an arbitrary observer on Earth's surface. A second Python code will produce an audio file with a series of wave forms in sync with the real-time acoustic simulations such that a noise resembling a passing shock wave will occur every time that the pressure front of a shock wave reaches the observer's position. This is to simulate the auditory effects that would be heard by an observer on Earth's surface.

Finally, the individual frames produced by each model are combined into a video file; for acoustic simulations, the audio file is added.

\section{\label{sec:sims}Simulation results}
We simulate the ground effects of a variety of hypothetical mitigation scenarios via fragmentation using PI. We investigate a wide range of parent asteroid diameters (20-800 m), with particular attention paid to size analogues of the 2013 Chelyabinsk asteroid (18 m; modeled here as 20 m) and the 1908 Tunguska impactor (modeled here as 50 m) as well as the size range of the initial, secondary, and final diameter estimates of the hypothetical 2023 PDC threat (200-800 m) \cite{chodas_2023_2023,wheeler_probabilistic_2023-1,wheeler_probabilistic_2023}. 

Each scenario presented below, with exception to the final 2023 PDC hypothetical impact scenario, assumes a parent asteroid with an average density of 2.6 $\text{g/cm}^3$ traveling at 20 km/s relative to Earth's reference frame with an entry angle of 45\textdegree relative to the horizon. All scenarios assume a spherical parent asteroid. Mitigated scenarios assume an average fragment disruption speed of $v_{disr}=1$ m/s.

We also model the ground effects of unmitigated (i.e., unfragmented) scenarios as a comparison. In the smaller threat regime (\textless100 m in diameter), it is possible that the parent asteroid, if left unmitigated, would airburst before making contact with the ground \cite{robertson_hydrocode_2019}.

We find that the ground effects of fragment airbursts from mitigation via PI are kept below their respective damage thresholds (optical energy deposition under 200 kJ/$\text{m}^2$ for each burst and sum of all shock wave over-pressures under 3 kPa at any given ground point) and are vastly lower than their unmitigated counterparts (Figures \ref{fig:small threats} and \ref{fig:large threats}, Tables \ref{table: data_table_frag} and \ref{table: data_table_unfrag}). To illustrate this, we plot the results of our ground effects simulations in the form of cumulative distribution functions (CDF) of both optical energy flux and acoustic over-pressure in the ground plane. We use the CDF of a particular threat scenario to determine whether that scenario results in ground effects of acceptably low magnitude. In addition to the maximum value, a useful metric is the 1$\%$ value of the CDF for a particular threat scenario, which is the magnitude at which 1$\%$ of ground locations resolved in the simulation experience optical energy flux or acoustic over-pressure above that value. We refer to this value henceforth as the 1$\%$ CDF value.

\subsection{\label{sec:small threats}Small threats (\textless100 m diameter)}
\begin{figure}[!ht]
    \centering
    \includegraphics[width=0.48\textwidth]{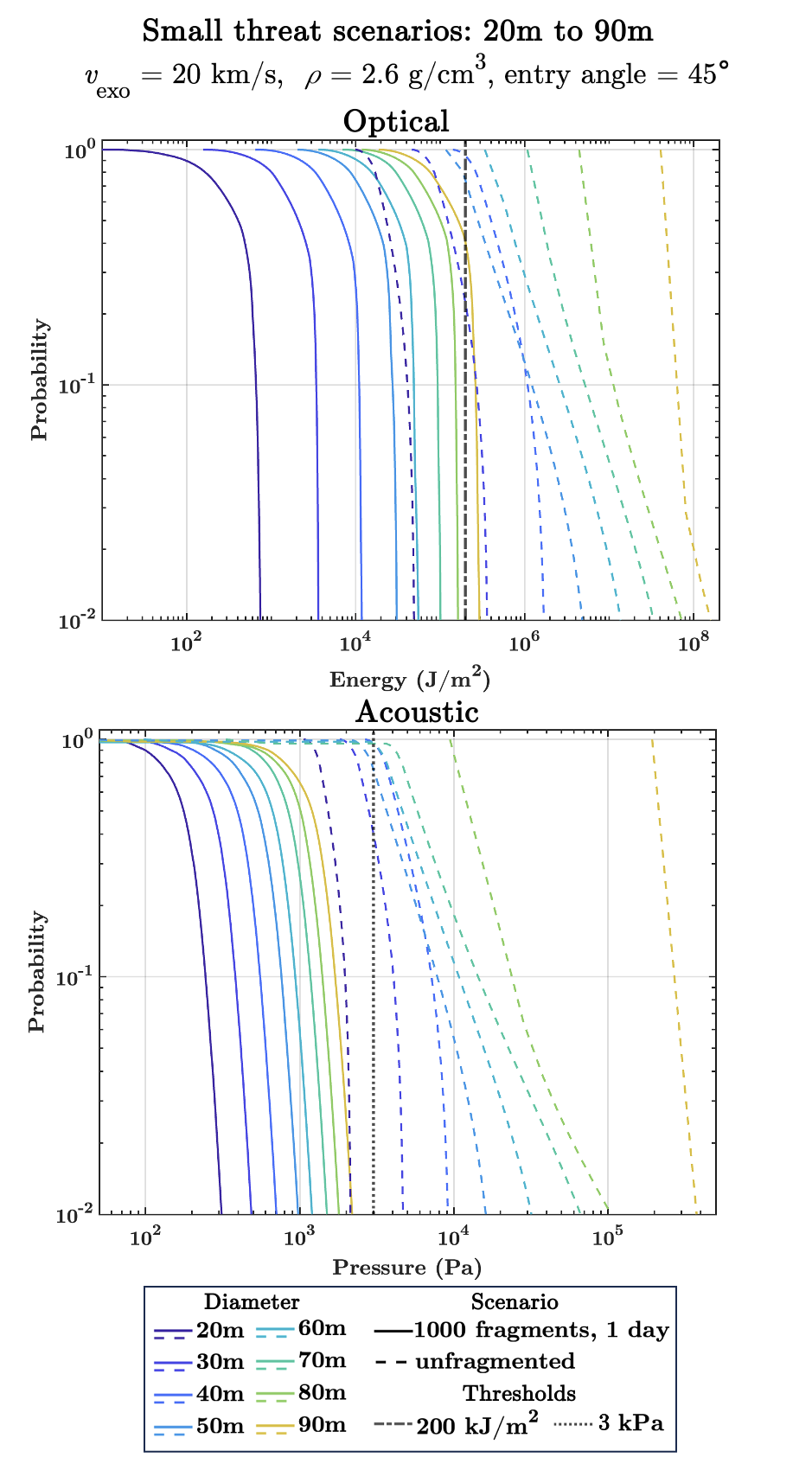}
    \caption{CDFs of the optical (upper) and acoustic (lower) ground effects for a variety of mitigation scenarios (solid lines) versus unmitigated scenarios (dashed lines) in the small threat regime (\textless100 m diameter). All scenarios assume a spherical parent asteroid with exo-atmospheric velocity ($v_{exo}$) of 20 km/s, average density ($\rho$) of 2.6 $\text{g/cm}^3$, and entry angle of 45\textdegree relative to Earth's horizon. Mitigated scenarios disassemble the parent asteroid into 1000 fragments with a one-day intercept prior to atmospheric entry and assume an average fragment disruption speed ($v_{disr}$) of 1 m/s. The dash-dotted black line (upper) and dotted black line (lower) mark the optical and acoustic damage thresholds of 200 kJ/$\text{m}^2$ and 3 kPa, respectively.}
    \label{fig:small threats}
\end{figure}

\begin{figure*}[!ht]
    \centering
    \includegraphics[width=0.8\textwidth]{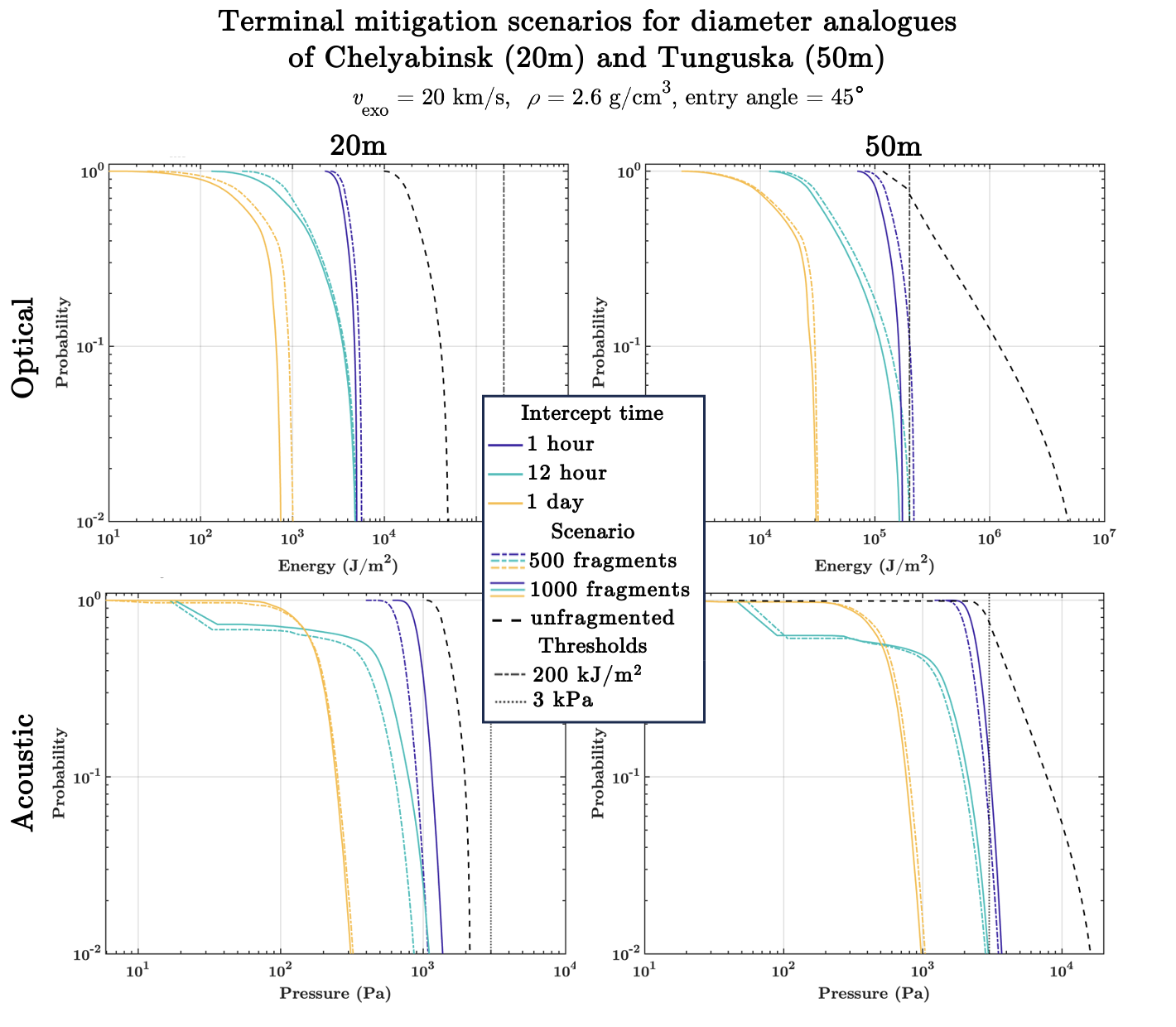}
    \caption{CDFs of the optical (upper) and acoustic (lower) ground effects for terminal mitigation scenarios of size analogues of the 2013 Chelyabinsk (20m) and 1908 Tunguska (50m) asteroids. All scenarios assume a spherical parent asteroid with exo-atmospheric velocity ($v_{exo}$) of 20 km/s, average density ($\rho$) of 2.6 $\text{g/cm}^3$, entry angle of 45\textdegree relative to Earth's horizon, and average fragment disruption speed ($v_{disr}$) of 1 m/s. The dash-dotted black line (upper) and dotted black line (lower) mark the optical and acoustic damage thresholds of 200 kJ/$\text{m}^2$ and 3 kPa, respectively.}
    \label{fig:20m 50m}
\end{figure*}

Airburst events within the small threat regime ($<$100 m diameter) can pose a significant hazard, as seen in historical cases like Chelyabinsk and Tunguska. Our simulations suggest that PI could effectively mitigate stony asteroids (average density of 2.6 $\text{g/cm}^3$) within the 20-90 m diameter range via disassembly into 1000 fragments with an intercept of $\sim$1 day prior to atmospheric entry, yielding ground effects that are much lower than from their unmitigated counterparts (Figure \ref{fig:small threats}, Tables \ref{table: data_table_frag} and \ref{table: data_table_unfrag}). The majority of these mitigated cases produce ground effects which are below their respective damage thresholds. The only exception to this is the optical output of the mitigated 90 m threat; however, the 1\% optical CDF value of the mitigated 90 m case is still $\sim$550 times smaller than that of an equal unfragmented case (Figure \ref{fig:20m 50m}, Tables \ref{table: data_table_frag} and \ref{table: data_table_unfrag}).

We highlight several terminal mitigation scenarios for stony asteroids with sizes similar to the Chelyabinsk and Tunguska asteroids at 20 m diameter and 50 m diameter, respectively. For each size, we simulate disassembly of the parent asteroid into 500 and 1000 fragments with intercept times (prior to atmospheric entry) of 1 hour, 12 hours, and 1 day to compare with an unfragmented airburst (Figure \ref{fig:20m 50m}, Tables \ref{table: data_table_frag} and \ref{table: data_table_unfrag}).

For the 20 m simulations, we find that all six scenarios are sufficient to keep all ground effects below their damage thresholds (Table \ref{table: data_table_frag}). The most conservative case of 500 fragments with an intercept of 1 hour yields 1\% CDF values of 5.7 kJ/$\text{m}^2$ for optical energy and 1.1 kPa for shock wave over-pressure, while a scenario of 1000 fragments with an intercept of 1 day results in 1\% CDF values of 0.8 kJ/$\text{m}^2$ and 0.3 kPa (Figure \ref{fig:20m 50m}, Table \ref{table: data_table_frag}). For comparison, we estimate that an equal unfragmented 20 m asteroid would yield an average optical energy deposition of 20.7 kJ/$\text{m}^2$ and average acoustic over-pressure of 1.4 kPa as experienced on the ground, with 1\% CDF values of 49 kJ/$\text{m}^2$ and 2.1 kPa, respectively (Figure \ref{fig:20m 50m}, Table \ref{table: data_table_unfrag}).

In the 50 m cases, most scenarios result in ground effects below the damage thresholds. Of the six scenarios investigated, the two cases with an intercept of 1 hour have the potential to exceed the thresholds. In the 500 fragment, 1 hour intercept scenario, we find that $\sim$10\% of ground locations are expected to exceed 200 kJ/$\text{m}^2$, while $\sim$5.5\% of ground locations exceed 3 kPa (Figure \ref{fig:20m 50m}, Table \ref{table: data_table_frag}). In the 1000 fragment, 1 hour intercept case, $\sim$13.5\% of ground locations experience shock wave over-pressures greater than 3 kPa (Figure \ref{fig:20m 50m}, Table \ref{table: data_table_frag}). Simulations of an unfragmented equivalent 50 m find an average optical energy deposition of 1044 kJ/$\text{m}^2$ and average shock wave over-pressure of 3.1 kPa, with 1\% CDF values of $\sim$4800 kJ/$\text{m}^2$ and 16.1 kPa (Figure \ref{fig:20m 50m}, Table \ref{table: data_table_unfrag}).

\subsection{\label{sec:large threats}Large threats (100-800 m diameter)}
We present mitigation scenarios for several large asteroids whose diameters span the size estimates of the hypothetical impact threat from the 2023 Planetary Defense Conference \cite{chodas_2023_2023,wheeler_probabilistic_2023-1,wheeler_probabilistic_2023}. Given the initial (April 2023) diameter estimate of 220-660 m and 1\% impact probability in October 2036 \cite{chodas_2023_2023}, an intercept mission designed for the worst-case scenario ($\geq660$ m diameter) could be prepared, and possibly launched, before the size of the asteroid is completely determined. For asteroids with diameters of 200-800 m (assuming a worst-case scenario up to $\sim125\%$ of the 660 m estimate), our simulations suggest that the PI method could effectively mitigate such threats with intercept times ranging from days to months (Figure \ref{fig:large threats}, Table \ref{table: data_table_frag}). 

We explore three scenarios spanning the initial estimated diameter range of 2023 PDC: 200 m, 350 m (similarly sized to 99942 Apophis), and 500 m (similarly sized to 101955 Bennu). We find that a 200 m asteroid broken into 30,000 fragments with a 10 day intercept yields 1\% CDF values of 43.9 kJ/$\text{m}^2$ and 1 kPa (Figure \ref{fig:large threats}, Table \ref{table: data_table_frag}). By disassembling an Apophis-sized threat into 50,000 fragments with a 30 day intercept, we achieve 1\% CDF values of 31.6 kJ/$\text{m}^2$ and 1.3 kPa (Figure \ref{fig:large threats}, Table \ref{table: data_table_frag}). A Bennu-sized threat broken into 100,000 fragments with a 60 day intercept results in 1\% CDF values of 37 kJ/$\text{m}^2$ and 1.5 kPa (Figure \ref{fig:large threats}, Table \ref{table: data_table_frag}). For comparison, our simulations suggest that each of these large diameter mitigation scenarios yields lower optical energy deposition and shock wave over-pressure than an unfragmented 20 m asteroid (Table \ref{table: data_table_frag}).

\subsection{\label{sec:800m}800 m 2023 PDC scenario}
\begin{figure}[!ht]
    \centering
    \includegraphics[width=0.48\textwidth]{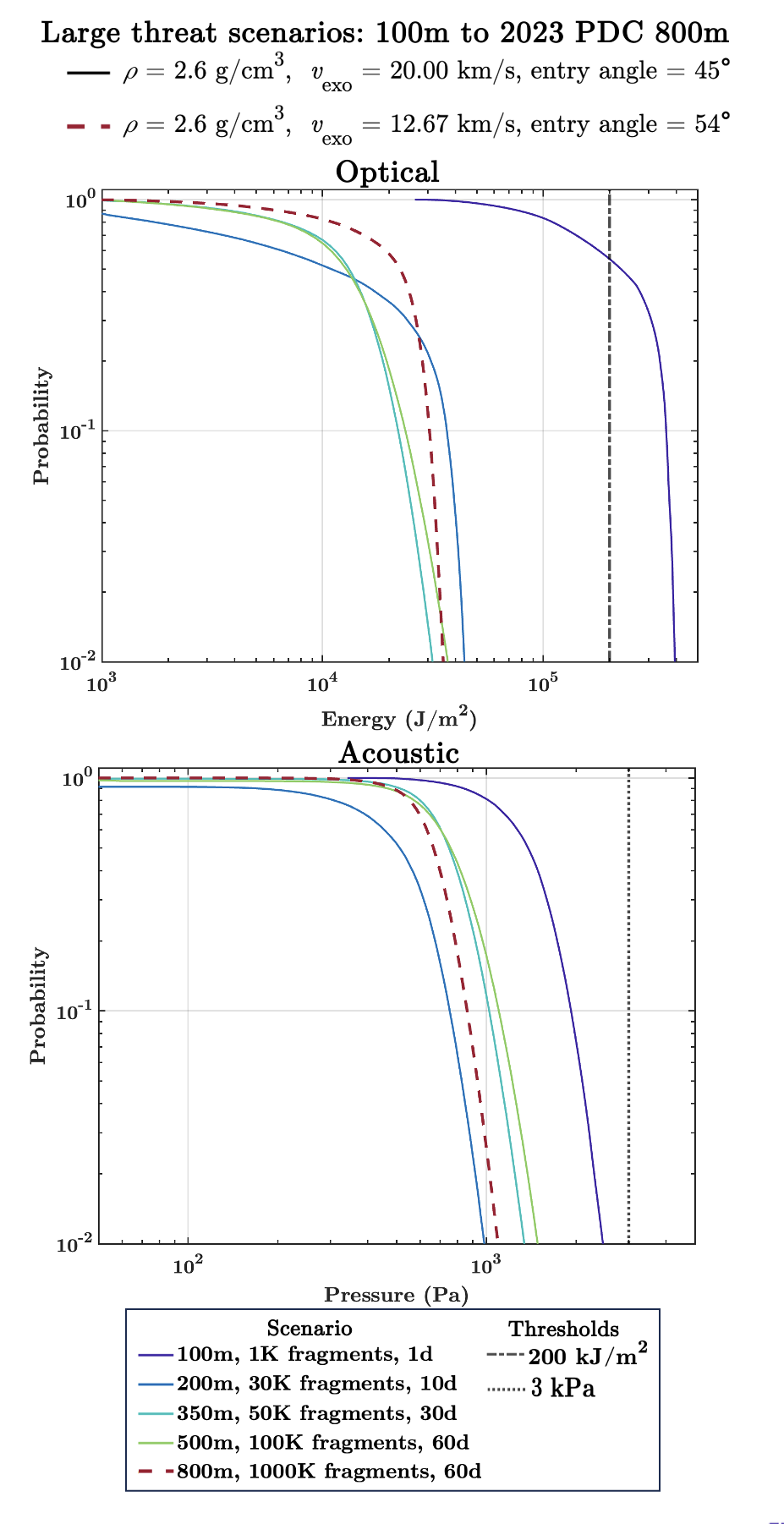}
    \caption{CDFs of the optical (upper) and acoustic (lower) ground effects for a variety of mitigation scenarios in the large threat regime (100-800 m diameter). All scenarios assume a spherical parent asteroid with an average density ($\rho$) of 2.6 $\text{g/cm}^3$ and average fragment disruption speed ($v_{disr}$) of 1 m/s. Solid lines dictate asteroids with exo-atmospheric velocity ($v_{exo}$) of 20 km/s and entry angle of 45\textdegree relative to Earth's horizon. The orange dashed line dictates a mitigation scenario for 2023 PDC, the hypothetical threat from the impact exercise of the 2023 Planetary Defense Conference. We utilize the final parameter measurements of 2023 PDC to model it as a stony asteroid (average density of $2.6$ $\text{g/cm}^3$) traveling at 12.67 km/s relative to Earth's reference frame with an entry angle of 54\textdegree \cite{wheeler_probabilistic_2023-1,wheeler_probabilistic_2023}. The dash-dotted black line (upper) and dotted black line (lower) mark the optical and acoustic damage thresholds of 200 kJ/$\text{m}^2$ and 3 kPa, respectively. }  
    \label{fig:large threats}
\end{figure}

\begin{figure*}[!ht]
    \centering
    \includegraphics[width=0.8\textwidth]{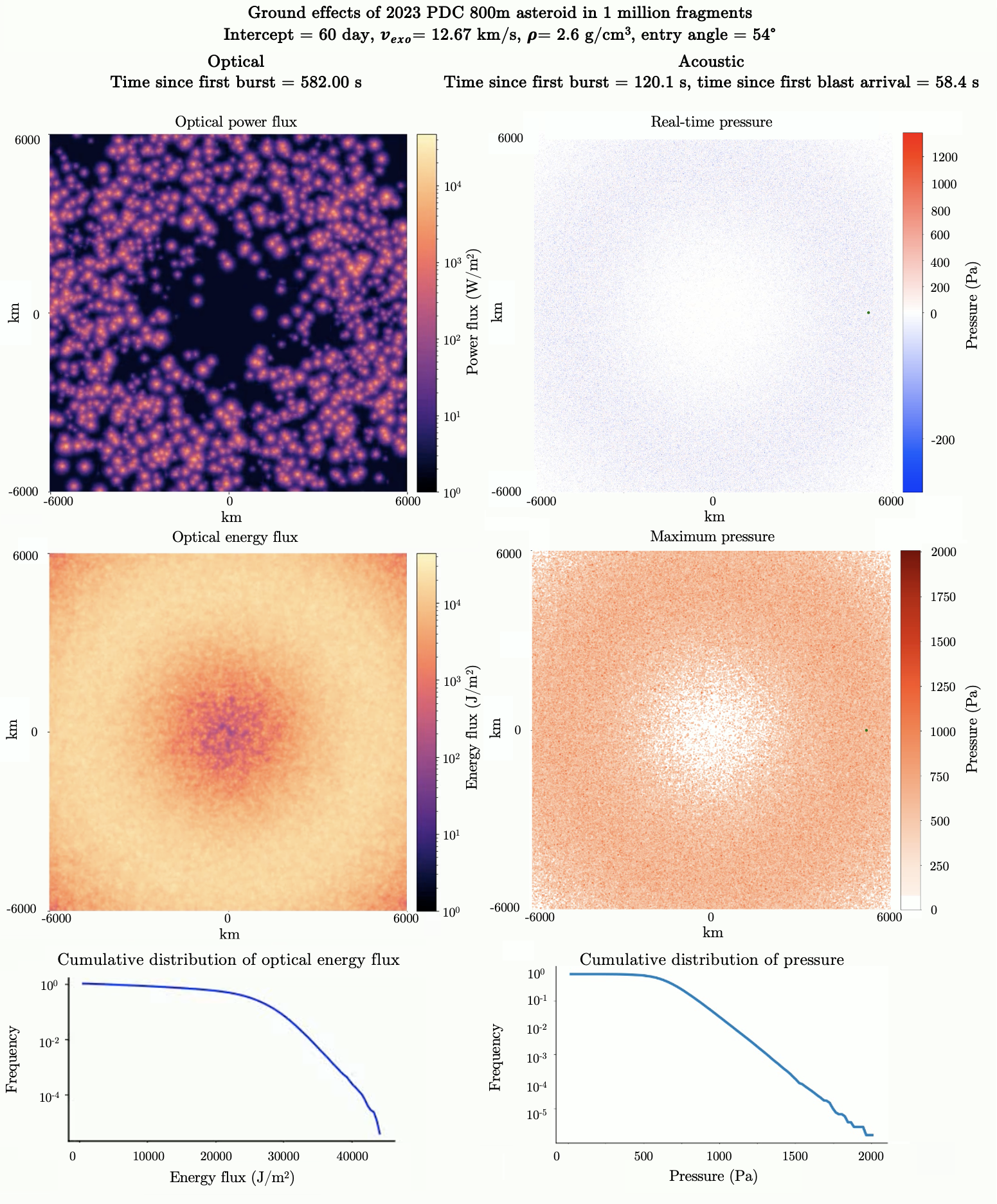}
    \caption{Optical (left) and acoustic (right) ground effects simulations showing a mitigation scenario of 2023 PDC, the hypothetical 800 m diameter threat from the impact exercise of the 2023 Planetary Defense Conference, broken into 1 million fragments with a 60 day intercept prior to impact. Simulations assume a spherical parent asteroid traveling at 12.67 km/s relative to Earth’s reference frame with an average density of 2.6 g/cm$^3$, entry angle of 54\textdegree, and average fragment disruption speed ($v_{disr}$) of 1 m/s. Note that the current time in each simulation differs; the real-time of each simulation is seen in the title as “Time since first burst” (units of seconds), dictating the amount of time that has passed since the first airburst of the first fragment. The acoustic simulation also includes the time since the first shock wave arrived at the observer (which is represented by the green dot in both plots). \textbf{Left:} optical pulse simulation. Upper: real-time optical power flux. Middle: real-time optical energy flux. Lower: CDF dictating the frequency of occurrence of various energy flux values. Note that the maximum optical energy deposition observed, which is experienced at $10^{-4}$\% of ground locations, is $\sim$4 times lower than the damage threshold of 200 kJ/m$^2$, and 1\% of observed ground locations experience 35.3 kJ/m$^2$. \textbf{Right:} acoustic shock wave simulation. Upper: real-time pressure. Middle: maximum pressure experienced in each location throughout the current length of the simulation; each pixel displays the highest pressure it has experienced. Dark orange planes show the caustics (the positive interference from interacting shock waves). Lower: CDF dictating the frequency of occurrence of various pressure values. Note that the sum of all shock wave over-pressures (including caustics) is lower than the damage threshold of 3 kPa; the maximum pressure observed, which is experienced at $10^{-4}$\% of ground locations, is 2 kPa, while 1\% of observed ground locations experience 1 kPa.}
    \label{fig:800m GE}
\end{figure*}

Simulations suggest that PI could effectively mitigate hypothetical asteroid 2023 PDC by disrupting the threat into 1 million fragments with a 60 day intercept prior to impact (Figures \ref{fig:large threats} and \ref{fig:800m GE}, Table \ref{table: data_table_frag}). We utilize the final parameter measurements of 2023 PDC to model it as a stony asteroid (average density of $2.6$ $\text{g/cm}^3$) traveling at 12.67 km/s relative to Earth's reference frame with an entry angle of 54\textdegree \cite{wheeler_probabilistic_2023-1,wheeler_probabilistic_2023}. 

In our mitigation scenario, we find that the asteroid's $\sim10.3$ Gt impact energy could be reduced to a series of distributed and de-correlated shock waves with an average acoustic over-pressure of $\sim$0.4 kPa as experienced by observers on Earth's surface (Figures \ref{fig:large threats} and \ref{fig:800m GE}, Table \ref{table: data_table_frag}). Including caustics, less than 1\% of locations on the ground experience pressures of $\sim1$ kPa, while less than $10^{-4}$\% of pressures reach $\sim2$ kPa (Figures \ref{fig:large threats} and \ref{fig:800m GE}, Table \ref{table: data_table_frag}). We find no locations in which a ground observer would experience any pressures at or above the 3 kPa damage threshold. We find an average optical energy deposition of 12.4 kJ/$\text{m}^2$, with 1\% of ground locations experiencing 35.3 kJ/$\text{m}^2$ (Figures \ref{fig:large threats} and \ref{fig:800m GE}, Table \ref{table: data_table_frag}). The maximum optical energy deposition observed, which is experienced at $10^{-4}$\% of ground locations, is $\sim$4 times lower than the damage threshold of 200 kJ/m$^2$ (Figures \ref{fig:large threats} and \ref{fig:800m GE}, Table \ref{table: data_table_frag}).

\section{\label{sec:conclusion}Conclusion}
Our simulations support the proposition that PI is an effective multi-modal approach for planetary defense that can operate in extremely short timescales as well as in extended modes to mitigate very large threats. Simulations in the terminal mode suggest that small stony asteroids (average density of 2.6 g/cm$^3$) of 20-90 m diameter can be mitigated via disassembly into 1000 fragments with an intercept of 1 day prior to atmospheric entry, with nearly all cases yielding ground effects below damage thresholds. We find that stony asteroids similarly sized to the Chelyabinsk ($\sim$20 m diameter) and Tunguska ($\sim$50 m diameter) impactors could be mitigated in terminal scenarios with a conservative number of fragments, with minimum feasible scenarios of disruption into 500 fragments with a 1 hour intercept for a 20 m threat and 1000 fragments with an intercept between 1-12 hours before impact for a 50 m threat.

By using PI in extended warning time scenarios, large stony asteroids with diameters of 100-800 m could be reduced to ground effects below their respective damage thresholds. Simulations suggest effective mitigation for an Apophis-sized threat (350 m diameter) via disruption into 50,000 fragments with a 30 day intercept; for a Bennu-sized threat (500 m), we find that 100,000 fragments with a 60 day intercept is feasible.

Simulations suggest that PI could effectively mitigate the hypothetical threat 2023 PDC from the impact exercise of the 2023 Planetary Defense Conference via disassembly into 1 million fragments with a 60 day intercept, yielding ground effects that are kept below established damage thresholds. We find that the asteroid's $\sim$10.3 Gt impact energy is reduced to in part to acoustic shock waves whose over-pressures are well below the 3 kPa damage threshold, with an average acoustic over-pressure of $\sim$0.4 kPa as experienced on the ground. Including cases where shock waves overlap (caustics), only 1\% of locations on the ground experience pressures of $\sim1$ kPa, while less than $10^{-4}$\% of pressures reach $\sim2$ kPa. We find no locations in which a ground observer would experience any pressures at or above the 3 kPa damage threshold; the maximum pressure observed is $\sim$1.5 times lower than the threshold. Optical ground effects simulations suggest an average optical energy deposition of of $\sim$12 kJ/$\text{m}^2$, with 1\% of ground locations experiencing $\sim$35 kJ/$\text{m}^2$. The maximum optical energy deposition observed is $\sim$4 times lower than the damage threshold of 200 kJ/m$^2$.

\section{Acknowledgments}
We gratefully acknowledge funding from NASA NIAC Phase I grant 80NSSC22K0764, NASA NIAC Phase II grant 80NSSC23K0966, and NASA California Space Grant NNX10AT93H, as well as from the Emmett and Gladys W. Fund. We gratefully acknowledge funding from NVIDIA for an Academic Hardware Grant for a high-end GPU which greatly speeds up ground effect simulations. Sandia National Laboratories is a multi-mission laboratory managed and operated by National Technology \& Engineering Solutions of Sandia, LLC (NTESS), a wholly owned subsidiary of Honeywell International Inc., for the U.S. Department of Energy’s National Nuclear Security Administration (DOE/NNSA) under contract DE-NA0003525. We would like to thank our large team of researchers and undergraduates at UCSB and beyond for their tireless work and enthusiasm to defend the planet.

\clearpage
\begin{table*}[!ht]
\caption{Summary of mitigated (fragmented) threat scenarios and estimated optical and acoustic ground effects. All scenarios assume a spherical parent asteroid with an average density ($\rho$) of 2.6 g/cm$^3$ and average fragment disruption speed ($v_{disr}$) of 1 m/s. All scenarios, with the exception of the 800 m 2023 PDC threat (case 23), assume an impact speed ($v_{exo}$) of 20 km/s and entry angle of 45\textdegree relative to Earth's horizon. Case 23 assumes an impact speed of 12.67 km/s and entry angle of 54\textdegree. \textit{D} indicates parent asteroid diameter. Intercept time is defined as the length of time prior to ground impact that the asteroid is intercepted.}

\centering
\begin{tblr}{width=1\linewidth,colspec={@{}X[c,valign=f]X[c,valign=f]X[c,valign=f]X[c,valign=f]X[c,valign=f]X[c,valign=f]X[c,valign=f]X[c,valign=f]X[c,valign=f]X[c,valign=f]@{}}
}
\hline
\hline
        Case no. & D (m) & No. fragments & Avg. fragment size (m) & Intercept time  & Unbroken exo-atm. energy (Mt) & 1\% optical CDF value (J/m$^2$) & Weighted avg. optical energy (J/m$^2$) & 1\% acoustical CDF value (Pa) & Weighted avg. pressure (Pa) \\ \hline
        1 & 20 & 500 & 2.52 & 1 hr & 5.21E-01 & 5.66E+03 & 3.59E+03 & 1.09E+03 & 5.87E+02 \\ 
        2 & 20 & 500 & 2.52 & 12 hr & 5.21E-01 & 4.98E+03 & 1.47E+03 & 8.61E+02 & 2.62E+02 \\ 
        3 & 20 & 500 & 2.52 & 1 d & 5.21E-01 & 1.01E+03 & 3.59E+02 & 3.23E+02 & 1.01E+02 \\ 
        4 & 20 & 1000 & 2.00 & 1 hr & 5.21E-01 & 5.02E+03 & 3.16E+03 & 1.37E+03 & 8.15E+02 \\ 
        5 & 20 & 1000 & 2.00 & 12 hr & 5.21E-01 & 4.79E+03 & 1.34E+03 & 1.10E+03 & 3.24E+02 \\ 
        6 & 20 & 1000 & 2.00 & 1 d & 5.21E-01 & 7.46E+02 & 2.57E+02 & 3.13E+02 & 9.80E+01 \\ 
        7 & 30 & 1000 & 3.00 & 1 d & 1.76E+00 & 3.63E+03 & 1.40E+03 & 4.87E+02 & 1.53E+02 \\ 
        8 & 40 & 1000 & 4.00 & 1 d & 4.16E+00 & 1.18E+04 & 4.60E+03 & 7.08E+02 & 2.21E+02 \\ 
        9 & 50 & 500 & 6.30 & 1 hr & 8.13E+00 & 2.18E+05 & 1.24E+05 & 3.50E+03 & 1.81E+03 \\ 
        10 & 50 & 500 & 6.30 & 12 hr & 8.13E+00 & 2.01E+05 & 5.76E+04 & 2.83E+03 & 8.26E+02 \\ 
        11 & 50 & 500 & 6.30 & 1 d & 8.13E+00 & 3.19E+04 & 1.27E+04 & 1.05E+03 & 3.21E+02 \\ 
        12 & 50 & 1000 & 5.00 & 1 hr & 8.13E+00 & 1.74E+05 & 1.03E+05 & 3.68E+03 & 2.06E+03 \\ 
        13 & 50 & 1000 & 5.00 & 12 hr & 8.13E+00 & 1.63E+05 & 4.83E+04 & 2.96E+03 & 8.84E+02 \\ 
        14 & 50 & 1000 & 5.00 & 1 d & 8.13E+00 & 3.09E+04 & 1.17E+04 & 9.77E+02 & 3.03E+02 \\ 
        15 & 60 & 1000 & 6.00 & 1 d & 1.41E+01 & 5.54E+04 & 2.11E+04 & 1.20E+03 & 3.75E+02 \\ 
        16 & 70 & 1000 & 7.00 & 1 d & 2.23E+01 & 1.01E+05 & 3.99E+04 & 1.49E+03 & 4.71E+02 \\ 
        17 & 80 & 1000 & 8.00 & 1 d & 3.33E+01 & 1.63E+05 & 6.59E+04 & 1.78E+03 & 5.63E+02 \\ 
        18 & 90 & 1000 & 9.00 & 1 d & 4.74E+01 & 2.92E+05 & 1.11E+05 & 2.20E+03 & 6.78E+02 \\ 
        19 & 100 & 1000 & 10.00 & 1 d & 6.51E+01 & 3.97E+05 & 1.53E+05 & 2.48E+03 & 9.37E+02 \\ 
        20 & 200 & 3E+04 & 6.44 & 10 d & 5.21E+02 & 4.39E+04 & 1.41E+04 & 9.93E+02 & 3.02E+02 \\ 
        21 & 350 & 5E+04 & 9.50 & 30 d & 2.79E+03 & 3.16E+04 & 8.74E+03 & 1.33E+03 & 4.17E+02 \\ 
        22 & 500 & 1E+05 & 10.77 & 60 d & 8.13E+03 & 3.70E+04 & 9.45E+03 & 1.47E+03 & 4.41E+02 \\ 
        23 & 800 & 1E+06 & 8.00 & 60 d & 3.33E+04 & 3.53E+04 & 1.24E+04 & 1.08E+03 & 3.58E+02 \\ \hline
\end{tblr}
\label{table: data_table_frag}
\end{table*}

\clearpage

\begin{table*}[!ht]
\caption{Summary of unmitigated (unfragmented) threat scenarios and estimated optical and acoustic ground effects. All scenarios assume a spherical parent asteroid with an average density ($\rho$) of 2.6 g/cm$^3$, impact speed ($v_{exo}$) of 20 km/s, and entry angle of 45\textdegree relative to Earth's horizon. \textit{D} indicates parent asteroid diameter.}
\centering
\begin{tblr}{width=1\linewidth,colspec={@{}X[c,valign=f]X[c,valign=f]X[c,valign=f]X[c,valign=f]X[c,valign=f]X[c,valign=f]X[c,valign=f]X[c,valign=f]X[c,valign=f]@{}}
}
\hline
\hline
        Case no. & D (m) & Unbroken exo-atm. energy (Mt) & 1\% optical CDF value (J/m$^2$) & Weighted avg. optical energy (J/m$^2$) & Max. optical energy observed (J/m$^2$) & 1\% acoustical CDF value (Pa) & Weighted avg. pressure (Pa) & Max. pressure observed (Pa) \\ \hline
        24 & 20 & 5.21E-01 & 4.90E+04 & 2.07E+04 & 4.98E+04 & 2.13E+03 & 1.38E+03 & 2.14E+03 \\ 
        25 & 30 & 1.76E+00 & 3.61E+05 & 1.25E+05 & 3.71E+05 & 4.67E+03 & 2.59E+03 & 4.72E+03 \\ 
        26 & 40 & 4.16E+00 & 1.68E+06 & 4.92E+05 & 1.78E+06 & 9.12E+03 & 4.23E+03 & 9.40E+03 \\ 
        27 & 50 & 8.13E+00 & 4.80E+06 & 1.04E+06 & 6.93E+06 & 1.61E+04 & 3.14E+03 & 1.95E+04 \\ 
        28 & 60 & 1.41E+01 & 1.38E+07 & 3.18E+06 & 2.49E+07 & 3.18E+04 & 5.48E+03 & 4.85E+04 \\ 
        29 & 70 & 2.23E+01 & 3.34E+07 & 9.36E+06 & 9.26E+07 & 6.65E+04 & 1.32E+04 & 1.65E+05 \\ 
        30 & 80 & 3.33E+01 & 7.54E+07 & 2.74E+07 & 4.18E+08 & 1.04E+05 & 4.74E+04 & 9.43E+05 \\ 
        31 & 90 & 4.74E+01 & 1.62E+08 & 9.69E+07 & 4.04E+09 & 3.87E+05 & 3.47E+05 & 1.93E+07 \\ \hline
\end{tblr}
\label{table: data_table_unfrag}
\end{table*}

\clearpage
\bibliographystyle{elsarticle-num-names} 
\bibliography{AA2023_PDC_acoustic-b}

\end{document}